# On artefact-free reconstruction of low-energy (30-250 eV) electron holograms

Tatiana Latychevskaia, Jean-Nicolas Longchamp, Conrad Escher, Hans-Werner Fink

Physics Institute, University of Zurich

Winterthurerstrasse 190, 8057 Zurich, Switzerland

Corresponding author: Tatiana Latychevskaia, tatiana@physik.uzh.ch

**Abstract**

Low-energy electrons (30-250 eV) have been successfully employed for imaging individual biomolecules. The most simple and elegant design of a low-energy electron microscope for imaging biomolecules is a lensless setup that operates in the holographic mode. In this work we address the problem associated with the reconstruction from the recorded holograms. We discuss the twin image problem intrinsic to inline holography and the problem of the so-called biprism-like effect specific to low-energy electrons. We demonstrate how the presence of the biprism-like effect can be efficiently identified and circumvented. The presented sideband filtering reconstruction method eliminates the twin image and allows for reconstruction despite the biprism-like effect, which we demonstrate on both, simulated and experimental examples.

1. Introduction

Low-energy electrons with kinetic energies in the range of 30-250 eV allow imaging individual biomolecules without detectable radiation damage [1]. A typical low-energy electron microscope is based on the original idea of holographic imaging proposed by Dennis Gabor [2-3]. It includes a source of coherent electrons, a sample placed a few microns from the source within the divergent electron wave and a detector unit positioned several centimetres away from the source [4]. The images recorded in a low-energy electron microscope are not direct images of objects but holograms that are formed by the interference of the scattered and non-scattered waves. By employing holographic imaging with low-energy electrons individual biological molecules such as DNA [5-7], the membrane protein rhodopsin [8], phthalocyaninato polysiloxane [9], the tobacco mosaic virus [10], a bacteriophage [11] and ferritin [12] were imaged. A good overview on imaging with low-energy electrons can be found in [13].

The plain setup of a low-energy electron microscope employs no lenses, and thus allows lens aberration-free imaging. However, it requires reconstruction of the objects from recorded holographic images which is not a trivial task. One artefact that plagues the holographic reconstructions is the twin image, which exists for all types of radiation and which was already discussed in Gabor's original work [2-3]. The twin image in inline holography is a second conjugated reconstructed image at double source-sample distance that is always superimposed on the reconstructed object. An experimental solution to the twin image problem was found introducing off-axis holography [14], which eventually could also be realized with low-energy electrons [15]. A numerical solution to the twin image problem is based on an iterative reconstruction process [16-17]. Another challenge associated with low-energy electron inline holographic imaging is specific to low-energy electrons. Previous studies demonstrated that when a freestanding object is positioned close to the source of coherent low-energy electrons, the trajectories of the electrons are deflected towards the object. Even the electrons that are not scattered by the object, and thus constitute part of the reference wave, slightly sheer towards the object [10, 18-19]. This effect is similar to the effect of electron deflection by a positively charged wire, and therefore often referred to as biprism-like effect [20]. This aberration of the reference wave complicates the reconstruction procedure as well as the interpretation of the retrieved structures. As a consequence, several groups replaced the reconstruction step by simulating holograms of positively charged objects that would fit experimental data [19, 21-22]. Hwang et al. [19] simulated holograms of carbon nanotubes and estimated that at relatively large source-sample distances, as for instance 11 μm at 340 eV, the biprism-like effect is negligible. But it becomes more pronounced at shorter distances. The latter however, are essential for imaging at higher magnification and hence, higher resolution. For example, at a source-sample distance of 1.8 μm (at 210 eV) a nanotube acts on the electron wave as a positively charged wire with 0.08 e/nm [19]. The biprism-like effect can be countered experimentally by placing the biomolecules onto a transparent support such as graphene [23-25]. The graphene support provides a plane at ground potential which helps to avoid high gradient fields around free-standing biomolecules and with that, the reference wave distortion [18]. The holograms of nano-objects placed on graphene can be easily reconstructed by applying conventional reconstruction routines based on classical optics equations [26]. Recently, it has been shown that the combination of inline holography and coherent diffractive imaging [27] when applied to images recorded with low-energy electrons comprises the potential of achieving atomic resolution in the reconstructed images [28].

In view of the perspective of low-energy electrons as non-destructive radiation for biological imaging at atomic resolution, the present work aims at obtaining artefact-free (i.e. neither twin image nor biprism-like effect) reconstruction of the structure of biomolecules from their holographic images by a non-iterative reconstruction routine.

## 2. Manifestation of the biprism-like effect in holograms

A molecule exposed to a low-energy electron beam can be perceived as a three-dimensional composition of individual scatteres (atoms), and the scattered wavefront at the detector plane is the superposition of wavefronts scattered by the atoms. The holographic pattern created by a single scatterer consists of concentric rings with decreasing periods towards the outer rings. Thus, the hologram of a molecule should appear as shadow projection with a superimposed interference pattern with decreasing fringe periods. However, it is often observed in low-energy electron holograms, that the interference pattern features equidistant fringes, resembling the interference pattern obtained when an electron biprism is introduced [20].

An electron biprism is created by placing a thin positively charged wire into the electron beam, which attracts the electrons towards the wire as they pass by. The electron wavefront is thus split into two parts that are bent towards each other and interfere behind the wire. The period of the equidistant fringes is related to the amount of charge on the wire and the wavelength $\lambda$ of the electrons.

The same interference pattern $I(x,y)$, which is created by an electron biprism, can be generated by two divergent spherical waves originating from two coherent point sources shifted aside $\pm x_0$ with respect to the real source:

$$I(x, y) = 2 + 2\cos\left(\frac{4\pi}{\lambda z} x x_0\right), \tag{1}$$

where $z$ is the distance between source plane and detector plane. Thus, the Fourier transform of the biprism interference pattern shows the two side peaks at $\nu = \pm 2x_0/\lambda z$, which defines the positions of the two virtual sources in the real source plane, as illustrated in Fig. 1. To verify any biprism-like distribution, the Fourier transform of the hologram must be studied.

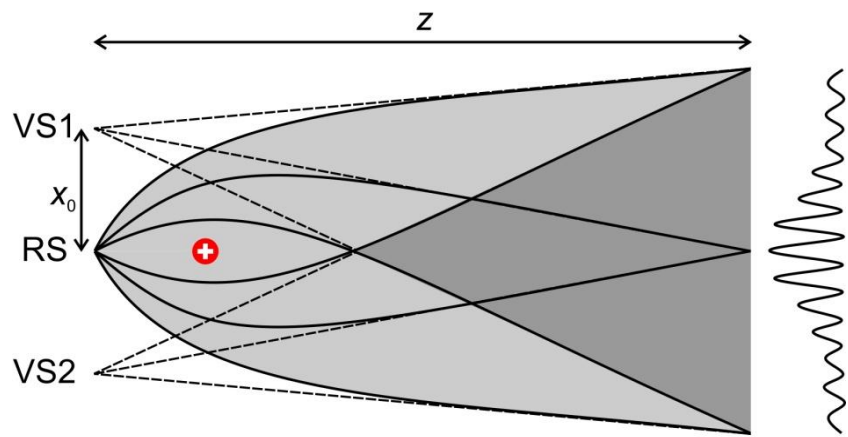

Fig. 1. Illustration of the electron trajectories around a positively charged wire leading to two virtual sources. RS represents the real source of electrons whereas VS1 and VS2 represent the virtual sources. The profile on the right side indicates the intensity distribution of the equidistant interference fringes.

Fig.2a shows an experimental low-energy electron hologram. The amplitude of the Fourier transform of the hologram exhibits three peaks which correspond to the zero frequency and two carrying frequencies of the equidistant fringe pattern, as demonstrated in Fig. 2b. It is therefore very easy to identify the presence of the biprism-like effect in the recorded hologram: its Fourier transform distribution has one main and two sidebands.

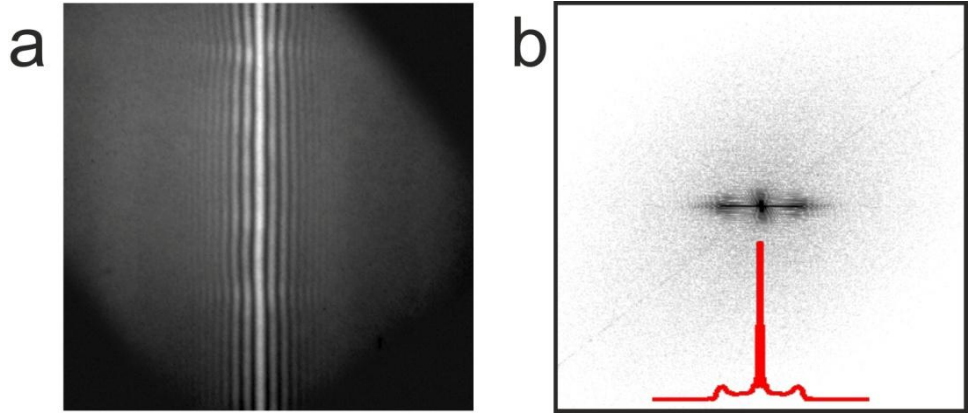

Fig. 2. Low-energy electron hologram of carbon nanotubes exhibiting the biprism-like effect. (a) Hologram recorded with 69 eV energy electrons, sampled with 1000×1000 pixels. (b) Amplitude of the Fourier transform of the hologram displayed on the logarithmic scale, only the central part (400×400 pixels) is shown. The red curve depicts the intensity profile at the centre of the image.

The presence of the three peaks in the Fourier transform of the hologram has been previously discussed by Georges et al [29], who recorded low-energy holograms of an intentionally charged tungsten tip. They could reconstruct the tip by introducing the reference wave in form of two divergent spherical waves originating from two ideal point-like sources whose positions were found from the Fourier transform of the hologram. However, for a more general holographic record this approach gets rather involved because the "charge distribution" follows the contour of the object, and is usually rather complex in the case of biological molecules. As a consequence, the reference wave cannot be determined easily, as the related reference wave sources are not point-like but rather form spread out spots, see the side peaks in Fig. 2b.

3. **Simulated holograms of charged objects**

Holograms affected by the biprism-like effect cannot be reconstructed by applying conventional reconstruction routines. While the object retrieved from biprism-free holograms appears in-focus, in holograms suffering from a biprism-like effect, the two reconstructions from the hologram formed by the two virtual sources overlap and obscure each other. It is therefore difficult to identify the correct in-focus distance and the object structure. In this section we discuss the simulation of holograms with biprism-like effect and in the next section we address the object reconstructions.

Quantitatively, the effect of a biprism, formed by a thin positively charged wire, on the electron trajectories can be estimated by using a cylindrical condenser model with a wire extended along $y$ direction and being at a potential $\varphi_0$ with respect to a grounded cylinder surrounding it. The potential distribution of the condenser as a function of the distance from the inner electrode $r$ is given by [20]:

$$\varphi(r) = \varphi_0 \frac{\ln(R_{out}/r)}{\ln(R_{out}/R_{in})}, \quad R_{in} \leq r \leq R_{out} \qquad (2)$$

$$r = \sqrt{x^2 + z^2}$$

The potential $\varphi_0$ can be replaced with the charge per unit length $Q_{lin}$:

$$\varphi_0 = \frac{Q_{lin}}{2\pi\varepsilon_0} \ln\left(\frac{R_{out}}{R_{in}}\right) \qquad (3)$$

and the electric field components are calculated as

$$\vec{E} = -\mathrm{grad}\,\varphi(r). \qquad (4)$$

When an electron is moving at a speed $v_0$ along z direction at a distance $r$ from the charged wire, it experiences the electric field component $E_x$ that deflects its trajectory:

$$E_x = \frac{Q_{lin}}{2\pi\varepsilon_0 r}. \qquad (5)$$

By solving the equations of motion for an electron, the angle of its deflection is found as:

$$\gamma = \frac{\Delta x}{\Delta z} \approx \frac{Q_{lin} e}{2\varepsilon_0 m v_0^2} \qquad (6)$$

and the introduced phase shift along x direction is therefore:

$$\Delta\chi = -\frac{2\pi}{\lambda}\gamma|x| = -\frac{\pi Q_{lin} e}{\lambda \varepsilon_0 m v_0^2}|x|. \qquad (7)$$

The total transmission function of an object located in the $(x_0, y_0)$ plane is described as:

$$T(x_0, y_0) = \exp(-a(x_0, y_0))\exp(i\Delta\chi(x_0, y_0)), \qquad (8)$$

where $a(x_0, y_0)$ is the absorption distribution of the object. The hologram of the object described by the transmission function $T(x_0, y_0)$ can be simulated by using conventional routines [27, 30-32].

To mimic an elongated biological molecule we simulated electron holograms of an object with the shape of Giacometti's figure "Standing Woman", as shown in Fig. 3. The object was assumed to be fully opaque. Holograms were simulated for electrons of 150 eV kinetic energy, the distance between the electron source and the sample was 1 μm, the distance between the source and the screen was 0.1 m and the images were sampled with 1000×1000 pixels. Holograms of the non-charged object ($Q_{lin}$=0 e/nm) as well as charged with $Q_{lin}$=0.1 e/nm and $Q_{lin}$=0.15 e/nm were simulated and are displayed in Fig. 3(b),(c) and (d), respectively. The slope of the resultant phase shift calculated with Eq.8 is 0.164 rad/nm for $Q_{lin}$=0.1 e/nm and 0.246 rad/nm $Q_{lin}$=0.15 e/nm.

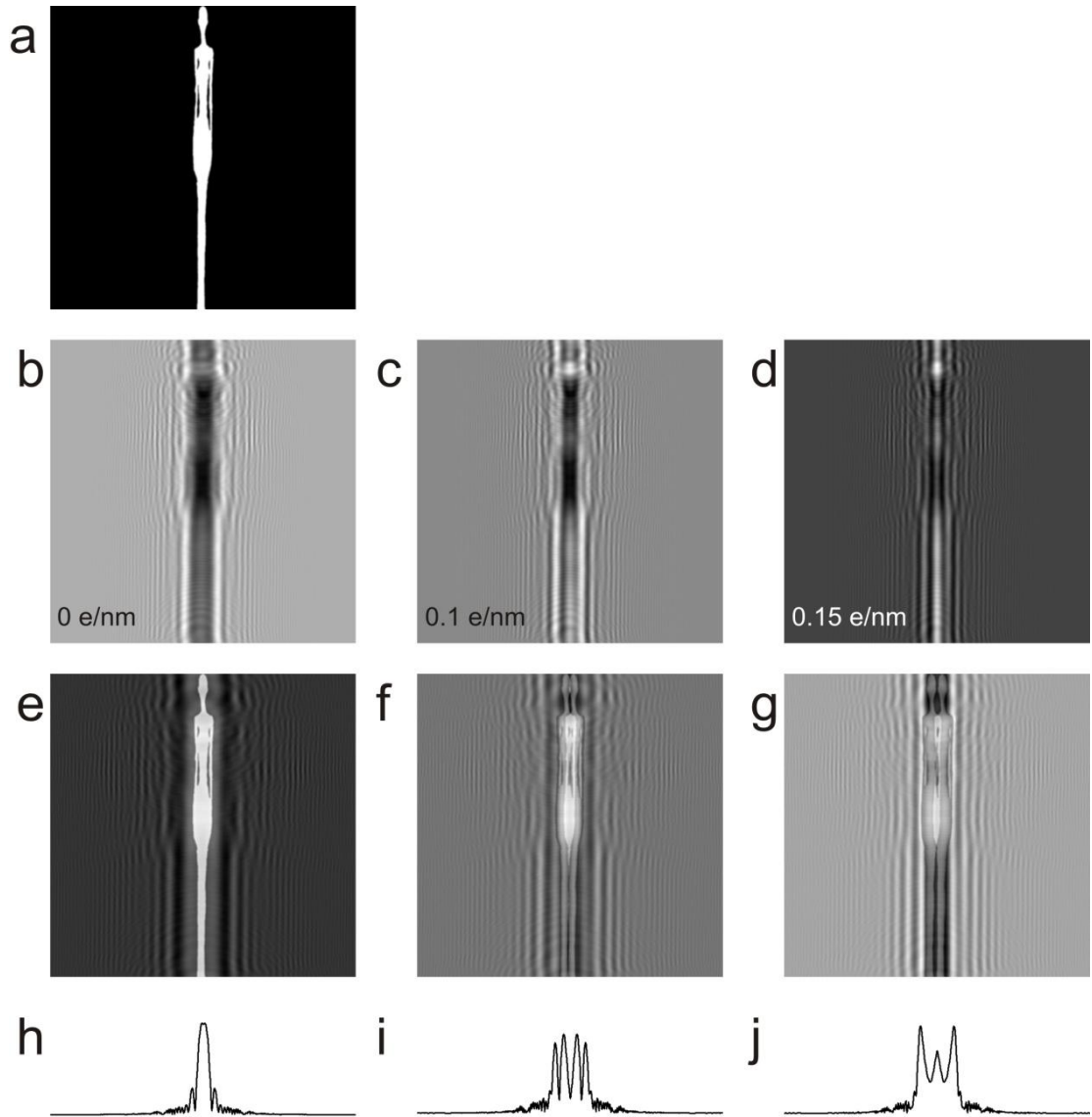

Fig. 3. Simulated hologram of a charged elongated object. (a) Absorption distribution of the object. (b)-(d) Simulated holograms of the object with different charge densities. (e)-(f) Corresponding reconstructions by conventional routines. (h)-(j) Amplitude profiles of the Fourier transform of the holograms shown for the central row of 500 pixels.

As can be seen from Fig. 3f-g, the conventional reconstructions of the holograms with biprism-like effect exhibit an overlap of two reconstructed objects. Especially, Fig. 3g represents an example of what might appear as an in-focus reconstruction if the correct shape of the object was not known beforehand. The amplitudes of the Fourier transformed holograms, shown in Fig. 3h-j demonstrate that in the absence of the biprism effect basically one central main peak is obtained, see Fig. 3h, whereas in the presence of the biprism effect pronounced side peaks developed, see Fig. 3i-j.

4. **Sideband filtering reconstruction**

The method of sideband filtered reconstruction of inline holograms was proposed in 1968 by O. Bryngdah and A. Lohmann [33]. It was an attempt to solve the twin image problem and it indeed allows elimination of the twin image, but only on one side of the object. The sideband filtering consists of taking the Fourier transform of a hologram, setting the left or the right half of the spectrum to zero, and taking the backward Fourier transform resulting in a filtered and generally complex-valued hologram. Subsequently, a conventional reconstruction routine is applied. In the resultant reconstruction, the twin image is eliminated on that side of the object where the spectrum was set to zero.

Applying the sideband filtering reconstruction to low-energy electron holograms not only selectively eliminates the twin image, but also allows overcoming the biprism-like effect for the following reason. The translation property of a Fourier transform implies that when a complex-valued function with linear phase distribution is Fourier transformed, its entire spectrum is shifted by a constant. Since the phase distribution imposed by a biprism is linear, see Eq.7, the Fourier transform of a hologram exhibits two sidebands surrounding the main peak. Both sidebands are identical Fourier spectra just shifted by a constant value defined by the strength of the biprism. Thus, three peaks in total are observed in the spectrum, as shown in Fig. 2b. By setting half of the Fourier spectrum of the hologram to zero, one sideband is eliminated. The reconstructed object is completely twin image-free on one side and has a well-defined contour on the other side. However, the entire reconstructed object is shifted according to the biprism strength, see Fig. 4a-f. Using the shift estimated from the Fourier spectrum and the reconstructed object contour, the left and right side can be shifted accordingly and recombined resulting in a qualitatively correct reconstruction, see Fig. 4g-i.

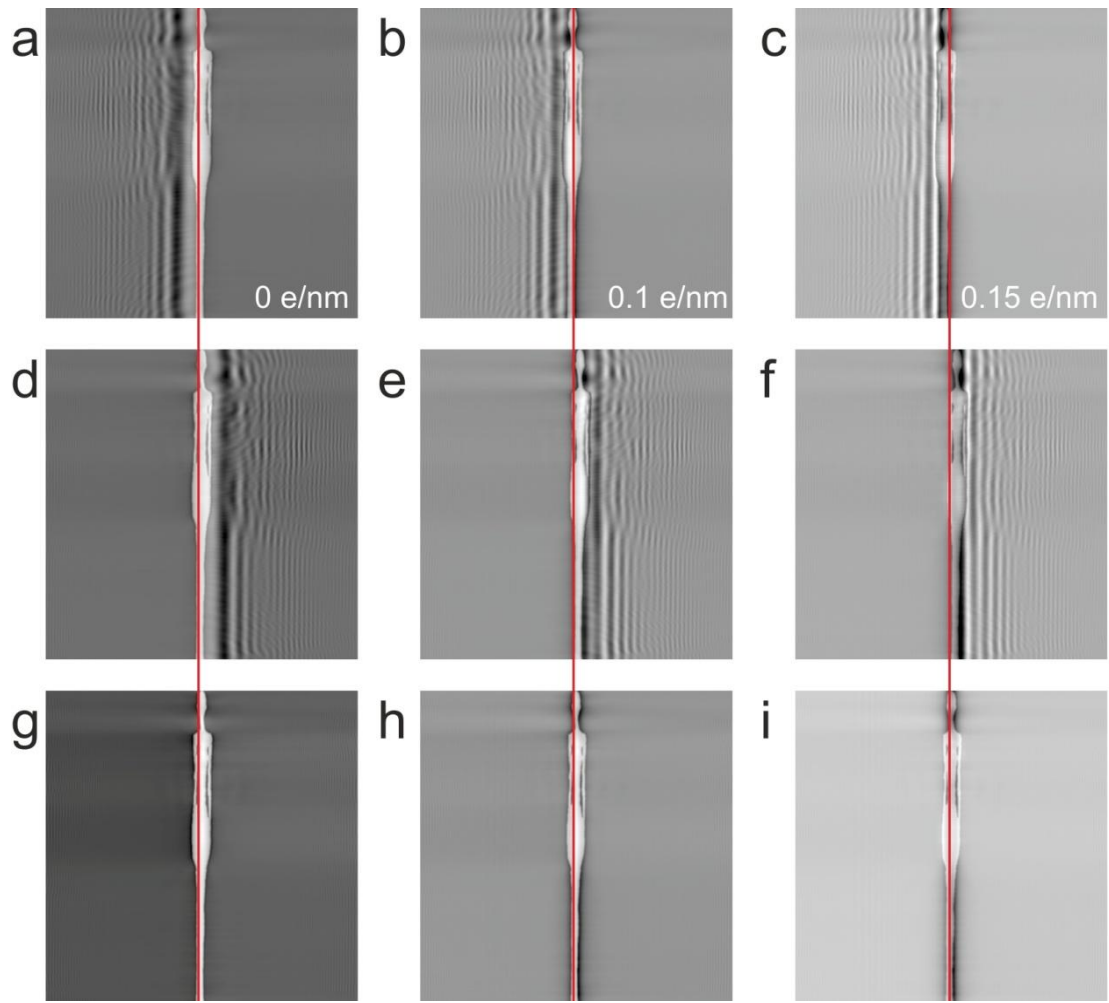

Fig.4 Sideband reconstructions of simulated holograms of a charged elongated object. Reconstructed absorption distribution when (a)-(c) the right side and (d)-(f) the left side of the Fourier spectrum is set to zero. (g)-(i) Recombined left and right twin image-free reconstructions of the object. The red lines help to assess the relative shifts of the reconstructions.

## 5. Experimental low-energy electron hologram

Of course, biomolecules are not always thin long objects extending along a certain direction. In general, they exhibit complex shapes which is an additional challenge for the reconstruction. In this section we show that the sideband filtering reconstruction can be applied to experimental data independently of the complexity of the geometrical shape of the object. The sample consists of a bundle of multi-walled carbon nanotubes suspended over a slit of about 200nm width milled in a thin carbon membrane by means of a focussed ion beam. The hologram was recorded with low-energy electrons of 100 eV kinetic energy. The recorded hologram and its conventional reconstruction, obtained by routines described elsewhere [27, 32], are shown in Fig. 5a and b, respectively. The reconstruction shown in Fig. 5b exhibits some object that might falsely be considered as regular (although twin-affiliated) reconstruction – a pitfall that we previously mentioned, compare Fig. 3g and Fig. 5b.

Next, we apply the sideband filtering reconstruction. Since the sample is a bundle of nanotubes which is attached to the edges of a carbon film, it cannot just be considered as a single rope-like object. Therefore, the sideband filtering cannot be done just by cutting the left or right side of the spectrum of the hologram. Instead, we applied a coordinate transformation to the hologram prior to sideband filtering. Since sideband filtering eliminates the twin image in the direction orthogonal to the interference fringes. First, the hologram is transformed into polar coordinates with the centre positioned in such a way that the radii are approximately perpendicular to the interference fringes, see the red dot and arrows in Fig. 5a. The result of this polar coordinate transformation is shown in Fig. 5c. 2. Thereafter, each row of the resultant image is subject to a one-dimensional Fourier transformation where upon the right side of the spectrum is set to zero. After a line by line backward Fourier transform, the obtained distribution is converted back to Cartesian coordinates. Finally, the resultant hologram is reconstructed, see Fig. 5d, where the right side of the sample free from the twin image is revealed. Analogical, the twin image-free reconstruction of the left side is obtained (Fig. 5e) and the two reconstructions are combined, as shown in Fig. 5f.

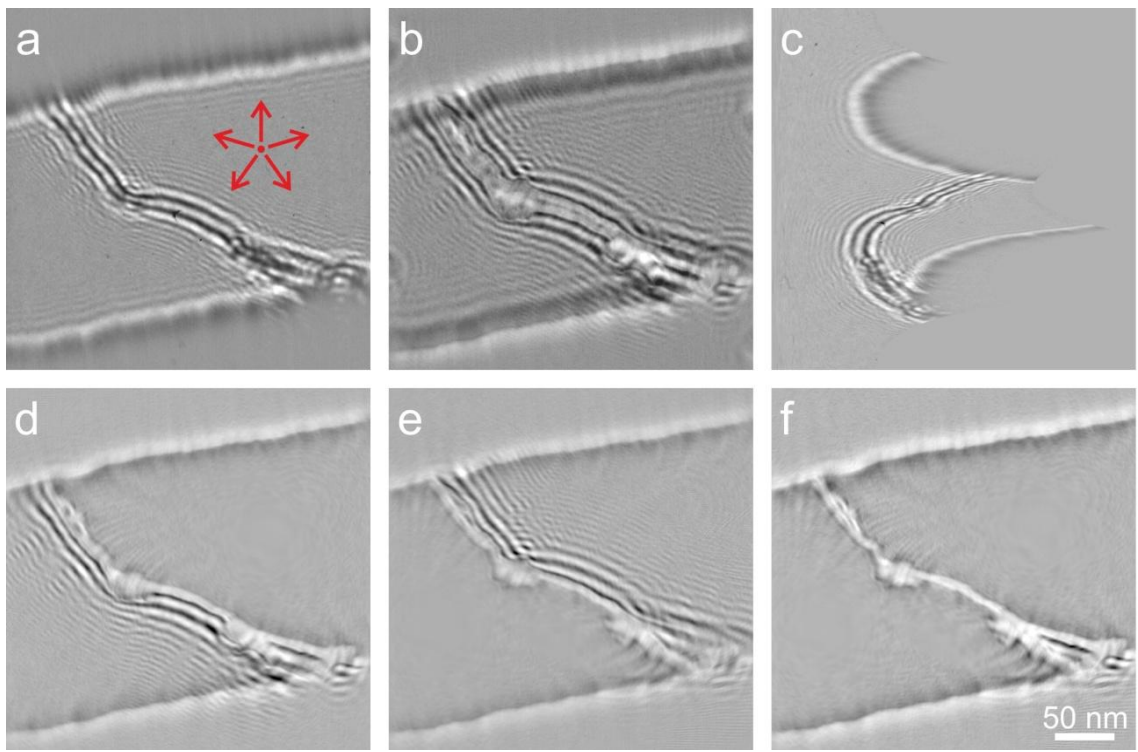

Fig.5 Sideband reconstruction of an experimental low-energy hologram. (a) Hologram of a bundle of multi-walled carbon nanotubes recorded with 100 eV electrons. The red dot and arrows indicate the position of the centre and directions of the transformation to polar coordinates for further sideband filtering. (b) Conventional reconstruction of the hologram. (c) Hologram transformed into polar coordinates. (d) Reconstruction of the right sideband filtered hologram. (e) Reconstruction of the left sideband filtered hologram. (f) Final reconstruction obtained by combining the right and left sideband filtered reconstructions.

## 6. Conclusions

Conventional reconstruction schemes do not lead to meaningful object representations once a biprism-like artefact is present. We provide a simple way to identify the presence of the biprism effect in holograms by examining the amplitude of the Fourier transform of a hologram and disclose a simple approach for artefact-free reconstruction of low-energy electron holograms. We show that the sideband filtering reconstruction method can be applied to circumvent the biprism effect and retrieve trustworthy twin-free structures of nanometer-sized objects, including individual biomolecules, from their holograms. In fact, we reconstruct simulated holograms of an extended object with a maximal linear charge of 1.5 e/nm sufficient to approximate the strongest biprism-like effect under typical experimental conditions. Artefact-free reconstructions from holographic records can serve as the initial step for further structure refinement by other high-resolution imaging techniques.

**Acknowledgements**

The authors are grateful for financial support by the Swiss National Science Foundation.